\documentclass[twocolumn,showpacs,amsmath,amssymb,amsfonts,floatfix,prl,aps]{revtex4}
\usepackage{graphicx}
\usepackage{color}

\begin{document}
\title{Bootstrap kernel for organic low dimensional systems; PPV, pentacene and picene}
\author{S. Sharma$^{1,2}$}
\email{sharma@mpi-halle.mpg.de}
\author{J. K. Dewhurst$^1$}
\author{E. K. U. Gross$^1$}
\affiliation{1 Max-Planck-Institut f\"ur Mikrostrukturphysik, Weinberg 2, 
D-06120 Halle, Germany.}
\affiliation{2 Indian Institute of Tecnology- Roorkee, Roorkee-247667, Uttarkhand, India.}
\date{\today}

\begin{abstract}
We apply the bootstrap kernel within time dependent density functional theory to study one-dimensional chain of 
organic polymer poly-phenylene-vinylene and molecular crystals of picene and pentacene. The behaviour of this kernel
in the presence and absence of local field effects is studied. The absorption spectra of poly-phenylene-vinylene has a 
bound excitonic peak which is well reproduced by the bootstrap kernel. Pentacene and picene, electronically similar 
materials, have remarkably different excitonic physics which is also captured properly by the bootstrap kernel. 
Inclusion of local-field effects dramatically change the spectra for both picene and pentacene. We highlight 
the reason behind this change. This also sheds light on the reasons behind the discrepancy in results between two different
previous Bethe-Salpeter calculations.  
\end{abstract}

\maketitle

Given the current urgency of investing renewable energy sources, it is difficult to overstate the importance of energy efficient systems like
organic solar cells and other opto-electronic materials.  The crucial information required for designing such systems is the optical absorption edge and 
spectra of the material, which in turn is dominated
by the physics of bound electron-hole pairs called excitons. In order to calculate accurate absorption spectra, in presence of excitons, one must
solve computationally expensive Bethe-Salpeter equation (BSE)\cite{sham66,hanke79}, which become even more cumbersome for systems of interest 
(with a few 100 atoms per unit-cell) for example in solar-cell technology. Time dependent density functional theory (TDDFT)\cite{tddft} is an alternative route
to calculate the exact absorption spectra with orders of magnitude less computational effort. 
However, within TDDFT the accuracy of the results rely entirely on the approximation used for the exchange-correlation (xc) kernel. 
In this regard, recently proposed bootstrap kernel\cite{sharma11} has shown very promising results for absorption spectra of 3D
periodic solids (II-IV and III-V insulators)\cite{sharma12}. However, it is not clear if the same bootstrap procedure can work for 
reduced dimensional systems which are of interest at present for energy efficient opto-electronics. 
In view of this we apply the bootstrap approximation to organic materials in one-dimension (poly-phenylene-vinylene (PPV) chain) and molecular crystals
(Picene and Pentacene). In the Later case the electronic properties are dominated by the electronic structure of the isolated molecules
representing a zero-dimensional case.

The choice of this set of materials is also motivated by the fact that each of these materials is well known to have 
rich excitonic physics\cite{draxl09,cudazzo12,ruini02} which is impossible to capture with simple xc kernels like 
random-phase approximation (RPA) or adiabatic local density approximation (ALDA)\cite{alda} and hence act 
as an ideal test bed for new xc kernels:\\
{\it PPV}-- 
organic conjugated polymers, of which PPV is an example, have novel opto-electronic properties which have been used
for the production of efficient light emitting diodes. One-dimensional chain of PPV  is one of the technologically 
important\cite{mukamel97,tian91,halliday93,alvarado98} conjugated polymer with strong electron-hole effects 
appearing as a bound excitonic peak 0.7eV below the fundamental gap\cite{mukamel97}. PPV chain thus acts as an ideal candidate to test any new xc 
kernel for the study of excitons in \emph{organic polymers} in general and in \emph{one-dimensional} systems in particular. 

{\it Picene and pentacene}-- 
usually, molecular solids are composed of molecules loosely bound to each other via van der Waals forces forming a crystal for which 
the electronic properties are still governed by individual molecules. This thus represents a zero-dimensional case for which
periodic boundary conditions are required. Such molecular solids are text-book examples of Frenkel excitons\cite{davydov,agranovich}, 
i.e. the electron-hole pair constituting an exciton are both localised on the same molecule. The situation becomes very interesting 
when the size of each molecular unit becomes large allowing for a possibility of the electron to be localised on one unit and the hole
hole on another forming a charge transfer exciton. Pentacene belongs to this Later class\cite{cudazzo12}. To make things
even more interesting, picene, which is structurally and electronically very similar to pentacene, has remarkably different optical 
properties dominated by Frenkel like excitons. 
Another important difference between the two is  that pentacene shows a strong Davydov splitting of the excitonic spectra while for 
picene this splitting is almost negligible\cite{dressel08,falter06}. These differences between the two, despite being very similar electronically, 
makes these systems interesting for studying the ability of the bootstrap kernel for capturing the subtle differences in the excitonic physics. 

The TDDFT equation for dielectric function is given by (atomic units are used):
\begin{align}\label{dyson}
 \varepsilon&^{-1}({\bf q},\omega)= 1+v({\bf q})\chi({\bf q},\omega)\\ 
 &=1+\chi_0({\bf q},\omega)v({\bf q})
 \left[ 1-(v({\bf q})+f_{\rm xc}({\bf q},\omega))\chi_0({\bf q},\omega)\right]^{-1} \nonumber
\end{align}
where $f_{\rm xc}({\bf q},\omega)$ is the xc kernel, $v$ is the bare Coulomb potential, $\chi$ is the full response function,
and $\chi_0$ is the response function of the non-interacting Kohn-Sham system.
The xc kernel in Eq. (\ref{dyson}) can be heurestically (but not uniquely) written as a sum of two terms, 
$f_{\rm xc}=f^{(1)}_{\rm xc}+f^{(2)}_{\rm xc}$. This partition of the kernel into two parts is done to capture two different effects -- 
(1) the band-gaps calculated using local/semilocal approximations to the xc potential within DFT are well known to be underestimated. 
In order to get the correct band structure one can perform a $GW$ calculation. Precicely the same effect can be obtained by the xc kernel 
without recourse to the many-body perturbation theory. $f^{(1)}_{\rm xc}$ is such a kernel and is responsible for 
correcting the underestimated band-gap, 
(2) the second part of the xc kernel, $f^{(2)}_{\rm xc}$, is responsible for capturing the excitonic physics. 
The Eq. (\ref{dyson}) can then be written as
\begin{align}\label{fxc2}
 \varepsilon&^{-1}({\bf q},\omega)= \\
 &1+\chi_{\rm gc}({\bf q},\omega)v({\bf q})
 \left[ 1-(v({\bf q})+f^{(2)}_{\rm xc}({\bf q},\omega))\chi_{\rm gc}({\bf q},\omega)\right]^{-1} \nonumber
\end{align}
where
\begin{align}\label{fxc1}
\chi_{\rm gc}({\bf q},\omega)=\left[ 1-\chi_{\rm 0} ({\bf q},\omega) f^{(1)}_{\rm xc}({\bf q},\omega)\right]^{-1}
\chi_{\rm 0}({\bf q},\omega),
\end{align}
is the gap corrected Kohn-Sham response function of the system. For all further calculations  we simply replace
$\chi_{\rm gc}$ by the response function calculated from the scissor operator corrected Kohn-Sham band structure-- unoccupied Kohn-Sham eigenvalues
are rigidly shifted to higher energies to make the Kohn-Sham and exact fundamental band-gap equal. In order to  keep the whole
procedure parameter free, the value of the fundamental band-gap is calculated using the $GW$ method\cite{cudazzo12,rholfing99,ruini02}.

The RPA is equivalent to setting $f^{(2)}_{\rm xc}=0$ and the bootstrap kernel\cite{sharma11} constitutes approximating $f^{(2)}_{\rm xc}$ by
\begin{align}\label{app}
 f^{\rm boot}_{\rm xc}({\bf q},\omega)=
 -\frac{\varepsilon^{-1}({\bf q},\omega=0)v({\bf q})}
 {\varepsilon_0^{00}({\bf q},\omega=0)-1}=
\frac{\varepsilon^{-1}({\bf q},\omega=0)}{\chi_{\rm gc}^{00}({\bf q},\omega=0)}
\end{align}
All these quantities (in Eqs. \ref{dyson}, \ref{fxc2}, \ref{fxc1} and \ref{app}) are matrices in the basis of reciprocal lattice 
vectors ${\bf G}$. $\chi^{00}$ represent 
the head of the susceptibility matrix  and $\varepsilon_0^{00}$ the head of the RPA dielectric tensor. 
Eqs. (\ref{app}) and (\ref{fxc2}) are solved self-consistently. 
The calculations are performed using the 
full-potentials localised augmented plane wave method\cite{singh} as implemented in the Elk code\cite{elk}. The {\bf k}-point mesh
required to achieve convergence is -- $6 \times 1 \times 1$ for PPV and $8 \times 6 \times 3$ for 
Picene and Pentacene.

The most expensive part of such a calculation is the matrix $\chi_0$. If one were to considers only the heads of the
matrices in Eqs. (\ref{dyson})-(\ref{app}) the calculations are exceptionally computationally efficient--
for pentacene calculations with head alone takes $<$1sec while the full matrix of 87 $\times$ 87 elements takes $\sim$1hour. 
Corresponding BSE calculation on the other hand require 25 hours on the same computer. Since the idea is to use TDDFT as a high throughput 
calculation for screening of thousands of possible energy efficient organic materials, we first check the quality of TDDFT 
calculations by using only the heads of the matrices in Eqs.(\ref{dyson})-(\ref{app}). 
Such a procedure is equivalent to ignoring local-field effects. The quality check for the final results can be performed
by comparing with full-BSE calculation and experimental data if available.


\begin{figure}[ht]
\includegraphics[width=\columnwidth]{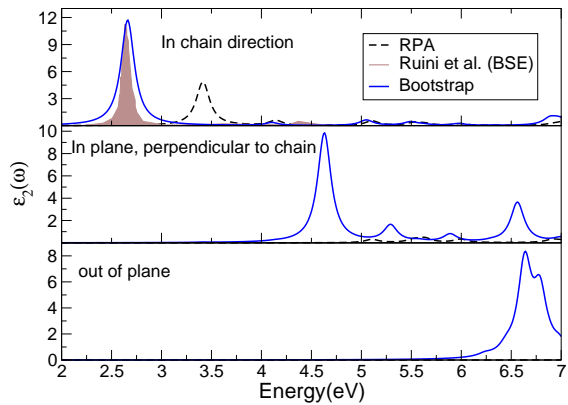} 
\caption{Dielectric tensor obtained using the bootstrap kernel as a function of energy in eV. 
The results obtained using the bootstrap kernel within TDDFT are shown with thick (blue) line. The RPA
results are shown with dashed(black) line. BSE data, shown as (brown) shaded area, is taken 
from Ref. \onlinecite{ruini02}. 
\label{ppv}}
\end{figure}

\emph{PPV}-- PPV is one of the conjugated polymer with 8 carbon atoms and 6 Hydrogen atoms per unit cell.
The results for the imaginary part of the dielectric function, $\varepsilon_2$, for PPV obtained with electric field polarized along the direction of the chain 
are presented in the top panel of Fig. (\ref{ppv}). 
The physics of bound electron-hole pair is, as expected, totally missing in the RPA results which shows a 
peak just above the quasiparticle gap\cite{rholfing99} at 3.3eV.
On the other hand, the results obtained using the bootstrap kernel show a strongly bound excitonic peak at 2.66eV which results in lowering 
of the absorption edge as compared to the RPA.
These results are in good agreement with the experimental absorption data\cite{mukamel97} which shows the first transition peak at 2.5eV.
It is also clear from Fig. \ref{ppv} that TDDFT results are in excellent agreement with the dielectric function obtained by solving the 
BSE \cite{ruini02,rholfing99}.
Other than the main peak at 2.5eV, the experimental absorption data\cite{mukamel97,halliday93} shows transitions at 3.7 and 4.8 and 6eV. 
These transitions are also very well reproduced by the bootstrap kernel showing peaks at 4.04, 5.05eV and 5.8-6eV. 

In the middle panel is shown the dielectric function obtained using the electric field in plane of the polymer but perpendicular 
to the chain and in the lower panel with electric field perpendicular to the polymer plane. In both cases the RPA spectrum is very 
different from the TDDFT spectrum which shows a large bound excitonic peak.

\begin{figure}[ht]
\includegraphics[width=\columnwidth]{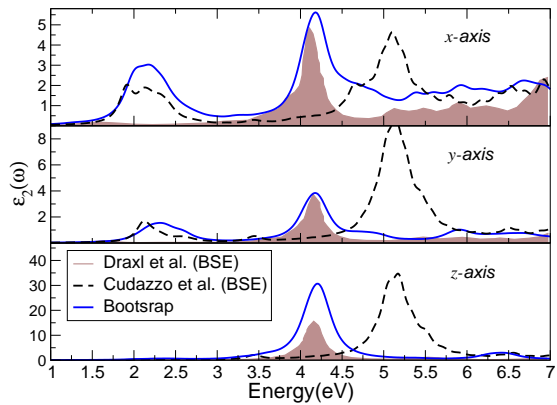} 
\caption{The dielectric tensor obtained using the bootstrap kernel  (thick (blue) line) as a function of energy in eV in three different 
polarization directions: $x$, $y$ and $z$. The BSE data of Draxl et al. is taken from Ref. \onlinecite{draxl09} and is shown as (brown) 
shaded area. Dashed (black) line is the BSE data of Cudazzo et al. from Ref. \onlinecite{cudazzo12}.
\label{pentacene-nlf}}
\end{figure}

\begin{figure}[ht]
\includegraphics[width=\columnwidth]{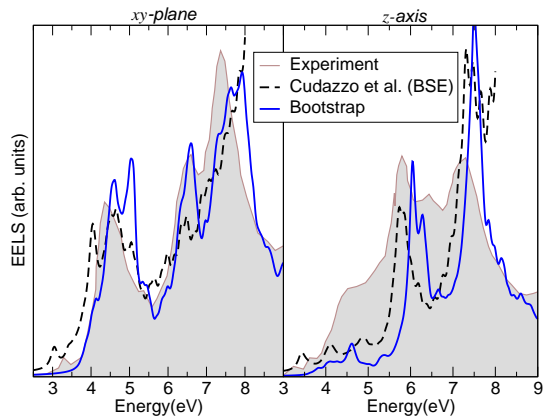} 
\caption{The electron energy loss spectra obtained using the bootstrap kernel (shown as thick (blue) line) as a function of energy in eV for 
two different directions: $xy$-plane (average of the response along $x$ and $y$-axis) and along $z$-axis. The experimental data 
is shown as (grey) shaded area and is taken from Ref. \onlinecite{roth10}. The BSE data, taken from Ref. \onlinecite{cudazzo12},
is shown as dashed (black) line. 
\label{picene-nlf}}
\end{figure}

Now we turn our attention towards the organic molecular crystals of picene and pentacene. The results for
three different polarization of light are shown in Figs. \ref{pentacene-nlf} and \ref{picene-nlf}. 
In order to compare with existing data, imaginary part of the dielectric function ($\varepsilon_2$) for pentacene and electron energy loss (EELS)
function for picene are presented. Both the materials are composed of five benzene rings joined in zigzag conformation in pentacene and in armchair
conformation in picene\cite{matt01,ghosh85}. There are two such units per primitive cell (for details of the structure see Fig. 1 of Ref. \onlinecite{cudazzo12}).
The Kohn-Sham gap obtained using LDA is small (2.45eV for picene and 0.68eV for pentacene) and a scissors correction is 
applied to make the Kohn-Sham gap equal to the quasiparticle gap which is 4.08eV for picene and 2.02eV 
pentacene. 

\emph{Pentacene}-- 
Due to low symmetry and molecular nature of the crystal the dielectric function is highly anisotropic. There are no transitions till 
the energy of 1.51eV for light polarized along $x$ and $y$-axis and till the energy of 3.5eV for the polarization axis along the $z$-axis. 
The main peak in the dielectric functions is at $\sim$4.1eV for all three polarizations of light. The first transition
peak is at 2.12eV for polarisation of light along $x$-axis and at 2.3eV for polarisation along $y$-axis. This shift in energy of the transition peak
along $x$ and $y$-axis is interpreted at the Davydov splitting\cite{cudazzo12,davydov}.
TDDFT results give a value of Davydov splitting to be 0.18eV, which is in excellent agreement with  the experimental data (0.15eV)\cite{dressel08,falter06}. 
As far as the comparison to BSE results is concerned, it is clear from Fig. \ref{pentacene-nlf} that, the TDDFT results 
calculated using the bootstrap kernel are in good agreement with the previous BSE results; in low energy region ($<$2.5eV) the results agree with 
the data of Cudazzo et al.\cite{cudazzo12} in terms of peak heights as well as peak positions. In high energy region ($<$3eV) the TDDFT 
results best reproduce the BSE data of Draxl et al.\cite{draxl09}. 

\emph{Picene}-- 
The EELS for solid picene is presented in Fig. \ref{picene-nlf}.
The TDDFT spectra in $xy$-plane (shown in the left panel) is in excellent agreement with the experimental data. we note that in terms of relative 
peak heights and positions the TDDFT results are in slightly better agreement with the experiments as compared to the BSE data (except for the initial 
peak at 3.3eV which is shifted to higher frequency in the TDDFT results). For the polarization vector along the $z$-axis the TDDFT and the BSE results
agree with each other but not with the experimental data in the low energy region ($<$5eV). In the energy range above 5eV agreement of 
calculations (both TDDFT and BSE) with the experiments is much better. It is interesting to note that, unlike pentacene, there is almost no Davydov 
splitting of the spectra in picene.

\begin{figure}[ht]
\includegraphics[width=\columnwidth]{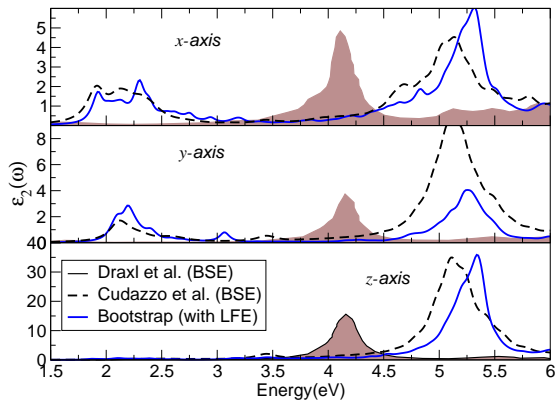} 
\caption{The dielectric tensor obtained using the bootstrap kernel (thick (blue) line) as a function of energy in eV in three different 
polarization directions: $x$, $y$ and $z$. These results are obtained by including local-field effects. 
The BSE data of Draxl et al. is taken from Ref. \onlinecite{draxl09} and is shown as (brown) 
shaded area. Dashed (black) line is the BSE data of Cudazzo et al. from Ref. \onlinecite{cudazzo12}. 
\label{pentacene-lf}}
\end{figure}
\begin{figure}[ht]
\includegraphics[width=\columnwidth]{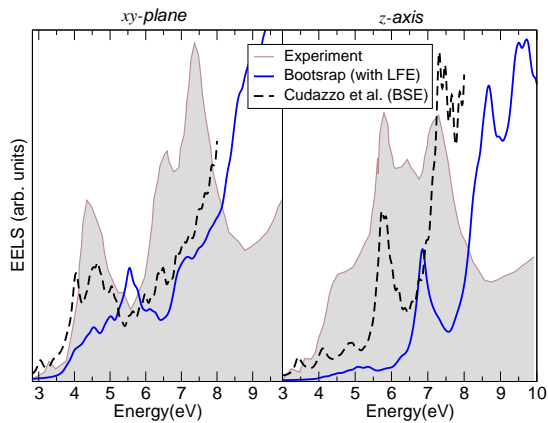} 
\caption{The electron energy loss spectra obtained using the bootstrap kernel (shown as thick (blue) line) as a function of energy in eV for 
two different directions: $xy$-plane (average of the response along $x$ and $y$-axis) and along $z$-axis. These results are obtained
by including local-field effects. The experimental data 
is shown as (grey) shaded area and is taken from Ref. \onlinecite{roth10} and the BSE data, taken from Ref. \onlinecite{cudazzo12},
is shown as dashed (black) line.
\label{picene-lf}}
\end{figure}
At this point it is important to ask the question; what happens to the spectra if the LFE are taken into account? To answer this in 
Figs. \ref{pentacene-lf} and \ref{picene-lf} are plotted the spectra for pentacene and picene including LFE. Both these materials are known to have 
large local field effects\cite{cudazzo12,draxl09}, inclusion of which makes the results for pentacene agree,
in both low and high energy range, with BSE data of Cudazzo et al.\cite{cudazzo12}. In order to obtain fully converged results in terms
of the size of the $v\chi_{\rm gc}$ we needed $87 \times 87$ matrix elements. These results seem to indicate that the  difference in the 
size of the screening matrix ($\epsilon_0 = 1+v\chi_{\rm gc}$) between the two BSE calculations (those of 
Cudazzo et al.\cite{cudazzo12} and Draxl et al.\cite{draxl09}) could be the reason for discrepancy between them.
In the case of picene the TDDFT results have the same shape as the BSE results but the peaks are shifted to higher energy by 0.85eV
and in general are in disagreement with experiments. 

In order to see what causes this large change in the spectra on inclusion of LFE  we analyzed the matrix $v\chi_{\rm gc}$ and found
the wings (i.e. elements with  either ${\bf G}=0$ or ${\bf G'}=0$) to be responsible. A closer look shows that 
the momentum matrix elements, 
required for the calculation of the wings, are very different in $x, y$ and $z$ 
directions due to the inhomogeneous environment in these molecular crystals and this difference is responsible for inducing large LFE.
As far as real 3D solids are concerned, a similar effect was also noticed by us for solid Ar, which being a noble gas solid is also strongly 
inhomogeneous.  
If one were to replace a momentum matrix element with its average (average of $x$, $y$ and $z$ direction), one obtains
TDDFT results (with LFE) for both picene and pentacene in close agreement with the BSE results. 
This is not a surprise because a similar average of the 
momentum matrix elements is used in the solution to the BSE to obtain the wings of the screening matrix ($\epsilon_0=1-v\chi_{\rm gc}$). 

It is clear from there results that bootstrap method in its simplest and computationally most efficient, i.e. just by using the head of the 
xc kernel, can be used for studying lower dimensional structures like conjugated polymers and  molecular crystals. The  difference 
in the physics of excitons between picene and pentacene is captured by the bootstrap kernel very well. Even subtle features like 
presence of small Davydov splitting in pentacene and absence in picene is well captured. The effect of inclusion of the local
fields is studied and we find that this shifts the
spectrum to higher frequency which in case of pentacene makes TDDFT results in very good agreement with one of the two previous BSE calculations.
The reason for large local-field effects was pinned down to the wings of the matrix of non-interacting response function times the Coulomb potential. 
This also elucidates the possible reason for discrepancy between the previous BSE results. Based on this insight we indicate a procedure, for
calculating the wings of the non-interacting response function, which makes TDDFT results mimic the BSE data for materials under investigation. 


\end{document}